\documentclass[prd,twocolumn,showpacs,amsmath,amssymb]{revtex4}
\usepackage{graphicx}
\usepackage{dcolumn}
\usepackage{bm}

\newcommand{\figwidth}{0.95\columnwidth}

\newcommand{\mpt}{\mbox{$\slash\kern-5.5pt p_T$}}
\newcommand{\mpx}{\mbox{$\slash\kern-5.5pt p_x$}}
\newcommand{\mpy}{\mbox{$\slash\kern-5.5pt p_y$}}

\begin{document}

\hspace{5.2in}\mbox{FERMILAB-PUB-07-553-E}

\title{Search for Randall-Sundrum gravitons with 1~fb$^{-1}$ of data from $p\overline{p}$ collisions at $\sqrt{s}=1.96$~TeV} 

%
\author{V.M.~Abazov$^{36}$}
\author{B.~Abbott$^{76}$}
\author{M.~Abolins$^{66}$}
\author{B.S.~Acharya$^{29}$}
\author{M.~Adams$^{52}$}
\author{T.~Adams$^{50}$}
\author{E.~Aguilo$^{6}$}
\author{S.H.~Ahn$^{31}$}
\author{M.~Ahsan$^{60}$}
\author{G.D.~Alexeev$^{36}$}
% ...

\affiliation{\vspace{0.1 in}(The D\O\ Collaboration)\vspace{0.1 in}}
% ...
\affiliation{$^{6}$University of Alberta, Edmonton, Alberta, Canada,
                Simon Fraser University, Burnaby, British Columbia, Canada,
                York University, Toronto, Ontario, Canada, and
                McGill University, Montreal, Quebec, Canada}
% ...
\affiliation{$^{29}$Tata Institute of Fundamental Research, Mumbai, India}
% ...
\affiliation{$^{31}$Korea Detector Laboratory, Korea University, Seoul, Korea}
% ...
\affiliation{$^{36}$Joint Institute for Nuclear Research, Dubna, Russia}
% ...
\affiliation{$^{50}$Florida State University, Tallahassee, Florida 32306, USA}
% ...
\affiliation{$^{52}$University of Illinois at Chicago,
                Chicago, Illinois 60607, USA}
% ...
\affiliation{$^{60}$Kansas State University, Manhattan, Kansas 66506, USA}
% ...
\affiliation{$^{66}$Michigan State University,
                East Lansing, Michigan 48824, USA}
% ...
\affiliation{$^{76}$University of Oklahoma, Norman, Oklahoma 73019, USA}
% ...

\date{October 17, 2007}
   
\begin{abstract}

Using 1 fb$^{-1}$ of data from $p\overline{p}$ collisions at $\sqrt{s}=1.96$~TeV at the Fermilab Tevatron collider collected by the D0 detector, we search for decays of Kaluza-Klein excitations of the graviton in the Randall-Sundrum model of extra dimensions to $e^+e^-$ and $\gamma\gamma$. We set 95\% confidence level upper limits on the production cross section times branching fraction which translate into lower limits on the mass of the lightest excitation between 300 and 900 GeV for values of the coupling $k/\overline{M}_{Pl}$ between 0.01 and 0.1.

\end{abstract}

\maketitle

The large difference between the Planck scale, $M_{Pl}\approx10^{16}$~TeV, and the weak scale presents a strong indication that the standard model is incomplete. In the presence of this hierarchy of scales it is not possible to stabilize the Higgs boson mass at the low values required by experimental data without an excessive amount of fine-tuning unless there is some, as yet unknown, physics at the TeV scale.

Randall and Sundrum have suggested a model~\cite{RS} in which the fundamental scale of gravity is near the weak scale and gravity appears so feeble because it is exponentially suppressed by the existence of a fifth dimension and a warped space-time metric. Standard model fields would be confined to one 3-brane (a 4-dimensional subspace of this 5-dimensional space) and gravity originates at another 3-brane. Only gravitons propagate in the bulk between these two branes. The apparent weakness of gravity originates from the small overlap of the graviton wave function with the standard model fields in the fifth dimension.

This model predicts a tower of Kaluza-Klein excitations as the 4-dimensional manifestation of the graviton propagating in 5-dimensional space. In the following we refer to these as RS (Randall-Sundrum) gravitons. The massless zero-mode couples with gravitational strength. The massive modes couple with similar strength as the weak interaction. Their properties are quantified by two parameters, the mass of the first massive excitation $M_1$ and the dimensionless coupling constant to standard model fields, $k/\overline{M}_{Pl}$, where $\overline{M}_{Pl}=M_{Pl}/\sqrt{8\pi}$ is the reduced Planck scale. To address the hierarchy problem without the need for fine-tuning $M_1$ should be in the TeV range and $0.01 < k/\overline{M}_{Pl} < 0.1$~\cite{DHR}. For these values the first massive RS graviton $G$ is a narrow resonance with a width much smaller than the resolution of the D0 detector. If kinematically accessible, RS gravitons can be resonantly produced in high energy particle collisions. They decay into pairs of fermions or bosons. 

In this Letter we consider decays into $e^+e^-$ and $\gamma\gamma$ pairs. We search for these as resonances in the $e^+e^-$ and $\gamma\gamma$ invariant mass spectrum from 1 fb$^{-1}$ of data collected using the D0 detector at the Fermilab Tevatron collider between October 2002 and February 2006. In the Tevatron protons and antiprotons collide at $\sqrt{s}=1.96$~TeV. D0 has previously published searches for RS gravitons~\cite{D0} and excluded $M_1< 250$~GeV for $k/\overline{M}_{Pl}=0.01$ and $M_1< 785$~GeV for $k/\overline{M}_{Pl}=0.1$ at 95\% confidence level with 260 pb$^{-1}$ of data. CDF has recently submitted for publication searches that exclude $M_1<889$~GeV for $k/\overline{M}_{Pl}=0.1$~\cite{CDF} based on 1.3 fb$^{-1}$ of data.

The D0 detector~\cite{run1det,run2det} consists of tracking detectors, calorimeters, and a muon spectrometer. The tracker employs silicon microstrips close to the beam and concentric cylinders of scintillating fibers in a 2~T axial magnetic field. The liquid-argon/uranium sampling calorimeter has an electromagnetic section that is 20 radiation lengths deep, backed up by a hadronic section. The calorimeter is divided into a central section covering $|\eta|\leq1.1$ and two endcap calorimeters extending coverage to $|\eta|\leq 4.2$. The luminosity is monitored by two arrays of plastic scintillation counters located on the inside faces of the endcap calorimeters. The pseudorapidity $\eta=-\ln[\tan(\theta/2)]$ and $\theta$ is the polar angle with respect to the proton beam direction. The azimuthal angle is denoted by $\phi$ and we measure object separation in the detector in terms of $\Delta R=\sqrt{(\Delta\phi)^2+(\Delta\eta)^2}$. We denote the momentum component transverse to the beam direction with $p_T$. Readout is controlled by a three-level trigger system. 

Since both electrons and photons result in electromagnetic showers with very similar signatures in our detector, we can define an inclusive selection that provides good efficiency for selecting $e^+e^-$ and $\gamma\gamma$ final states. In particular we require clusters of energy depositions in the electromagnetic calorimeter that are consistent with the expected shower profile using a $\chi^2$ test and have less than 3\% of their energy leaking into the hadronic calorimeter section. We require that the cluster is well isolated with less than 7\% of the cluster energy in an annular isolation cone with $0.2 < \Delta R < 0.4$ around the cluster centroid and less than 2~GeV for the sum of the $p_T$ of all tracks with $0.05 < \Delta R < 0.4$ with respect to the cluster centroid. To accept both electrons and photons we do not require a matched track. We start with a data set of 34 million events triggered on one or two electromagnetic showers with $p_T$ thresholds between 15 and 35 GeV. We select events in which there are at least two such clusters with $p_T>25$~GeV in the central calorimeter with $|\eta|<1.1$. Including clusters in the end calorimeters would add little acceptance for decay products of massive objects. In the collider data we find 43639 events that satisfy these selection criteria with the invariant mass of the two clusters $M_{ee/\gamma\gamma}>60$~GeV.

\begin{figure}[htph]
\includegraphics[width=\figwidth]{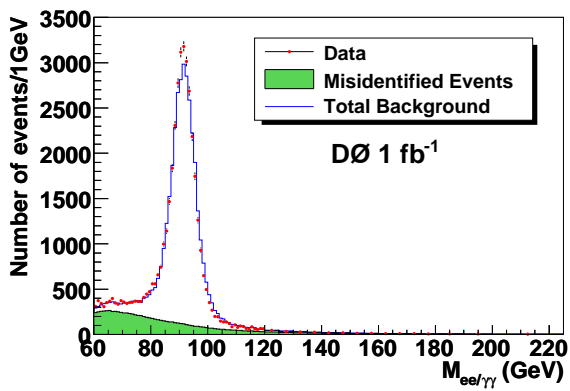}
\caption{\label{fig:lowmass} Invariant mass spectrum from data (points). Superimposed is the fitted total background shape from standard model processes including events with misidentified clusters (open histogram) and the fitted contribution from events with misidentified clusters alone (shaded histogram).}
\end{figure}

Within the standard model, the Drell-Yan process and diphoton production give rise to $e^+e^-$ and $\gamma\gamma$ final states. The invariant mass spectrum for these is expected to fall towards higher masses except for the $Z\to e^+e^-$ resonance. We model these backgrounds using a Monte Carlo simulation with the {\sc pythia}~\cite{pythia} event generator using the CTEQ6L parton distribution functions~\cite{CTEQ6}, followed by a {\sc geant}-based~\cite{geant} detector simulation. Another source of events is the misidentification of one or two jets as electron or photon candidates. The shape of the invariant mass spectrum of this source of events is estimated from data by selecting events with energy clusters in the electromagnetic calorimeter that are not consistent with electromagnetic showers and fail the $\chi^2$ test for the shower profile. The absence of the $Z$ resonance in the background spectrum in Fig. \ref{fig:lowmass} confirms that this sample has no significant contamination from $e^+e^-$ final states.

We fit the shape of the invariant mass spectrum from the data near the $Z$ resonance ($60 < M_{ee/\gamma\gamma} < 140$~GeV) with a superposition of the spectrum from Monte Carlo predictions for the standard model processes and the spectrum expected from misidentified clusters. In the fit, the spectra from $e^+e^-$ and $\gamma\gamma$ final states are normalized relative to each other by the leading order cross section from {\sc pythia}, the total number of events is fixed to the number of events observed in the data, and the fraction $f$ of all events that have misidentified clusters is the only free parameter. We obtain best agreement with the data for $f=0.21\pm0.01$. The spectra are shown in Figure~\ref{fig:lowmass}. Trigger thresholds affect the shapes near the low mass end of the fit window. We account for this by assigning a systematic uncertainty on the value of $f$. At masses above 100~GeV the trigger is fully efficient. 

\begin{figure}[htph]
\includegraphics[width=\figwidth]{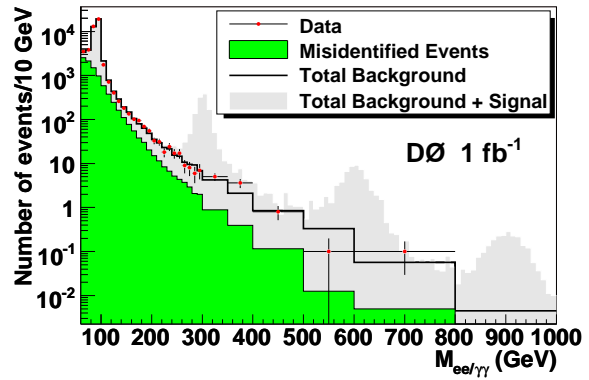}
\caption{\label{fig:allmass} Invariant mass spectrum from data (points). Superimposed is the fitted total background shape from standard model processes including events with misidentified clusters (open histogram) and the fitted contribution from events with misidentified clusters alone (shaded histogram). The grey shaded histogram shows the signal expected from gravitons with $M_1=300$, 600, and 900~GeV and $k/\overline{M}_{Pl}=0.1$ on top of the total background.}
\end{figure}

We compare the invariant mass spectrum of our background model with the fitted value of $f$ to the data at higher masses. As shown in Figure~\ref{fig:allmass}, we find agreement between background model and data in the high-mass range. There is a slight mismatch in the mass resolution at the $Z$ peak between our Monte Carlo simulation and the data. We verified that this does not affect the predictions of the background model at higher masses. 

From the fitted number of $p\overline{p}\to e^+e^-+X$ events (most of them in the $Z$ resonance), the acceptance and efficiency from the Monte Carlo simulation, and the calculated standard model cross section, we determine the integrated luminosity of the data sample. All Monte-Carlo derived efficiencies are multiplied by 0.96 so that the efficiency from the $Z\to e^+e^-$ Monte Carlo simulation agrees with the efficiencies measured in $Z\to e^+e^-$ data. The leading order cross section for the $e^+e^-$ final state with $60<M_{ee}<130$~GeV from {\sc pythia} is 178 pb. We multiply this by a next-to-leading order (NLO) $K$-factor of 1.34~\cite{Z}. This gives $985\pm35$~pb$^{-1}$. The uncertainty in this number is dominated by the uncertainty in the cross section from parton distribution functions. We do not include uncertainties on efficiencies and acceptances because these cancel in the limit calculation. This value is in agreement with the number determined using the luminosity counters ($1036\pm63$~pb$^{-1}$)~\cite{D0lumi}. 

We determine the signal acceptance and efficiency using a Monte Carlo simulation of RS gravitons with $200<M_1<1000$~GeV using {\sc pythia} and {\sc geant}. Systematic uncertainties in the signal efficiency originate from detector resolution (1-11\%), parton distribution functions (0.2-5.5\%), electron and photon identification efficiencies (1.4\%), and the finite signal Monte Carlo sample size (0.5\%). Contributions to the uncertainty in the background prediction are from the finite size of Monte Carlo and data samples (2-24\%), parton distribution functions (2-10\%), the mass dependence of the NLO $K$-factor (5\%), and the uncertainty in the trigger thresholds (1\%). In some cases the uncertainties vary with the invariant mass value.

\begin{figure}[btp]
\includegraphics[width=\figwidth]{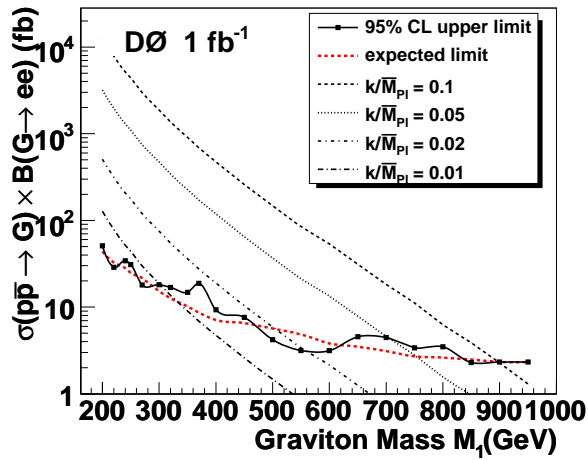}
\caption{95$\%$ confidence level upper limit on $\sigma(p\overline{p}\rightarrow G+X)\times B(G\rightarrow e^+e^-)$ from $~$1 fb$^{-1}$ of data compared with the expected limit and the theoretical predictions for different couplings $k/\overline{M}_{Pl}$.}
\label{fig:xseclimit}
\end{figure}

We compute upper limits for the production cross section of RS gravitons times branching fraction into $e^+e^-$ final states at 95\% confidence level by comparing the observed and expected numbers of events in a sliding mass window. The width of the window was optimized for maximum sensitivity using the Monte Carlo simulation and varies from 20 GeV for $M_1=200$~GeV to 120~GeV for $M_1=950$~GeV. We use a Bayesian approach to integrate over all important input parameters such as signal efficiency, background prediction, and integrated luminosity, using a Gaussian prior with width equal to the estimated uncertainties in the parameters~\cite{D0limit}. For the RS graviton production cross section we use a flat prior. To compute the limits, we use the integrated luminosity determined from the $Z$ signal, which gives us a more precise normalization than the direct luminosity measurement. Figure~\ref{fig:xseclimit} shows the limits as a function of invariant mass compared to predictions from the Randall-Sundrum model and Table~\ref{tab:limits} tabulates the results. Based on the observed and expected numbers of events we obtain limits on $\sigma(p\overline{p}\rightarrow G+X)\times B(G\rightarrow e^+e^-/\gamma\gamma)$. We divide by $B(G\rightarrow e^+e^-/\gamma\gamma)/B(G\rightarrow e^+e^-)=3$~\cite{BR} to convert these to the quoted limits on $\sigma(p\overline{p}\rightarrow G+X)\times B(G\rightarrow e^+e^-)$.

\begin{table*}
\begin{center}
\caption{Input data for limit calculation and 95\% confidence level limits on cross section times branching fraction and coupling. Quoted are the total uncertainties that are used in the limit calculation.}
\label{tab:limits}
\begin{tabular}{@{\extracolsep{0.15in}}cccccccccc}
\hline\hline
      &        &      &       &        & \multicolumn{3}{c}{$\sigma(p\overline{p}\rightarrow G+X)\times B(G\rightarrow e^+e^-)$ (fb)} & \multicolumn{2}{c}{$k/\overline{M}_{Pl}$}\\
$M_1$ & window & data & back- & signal & theory & expected & observed & expected & observed \\
(GeV) & (GeV) & & ground & efficiency & & limit & limit & limit & limit\\
\hline\hline
200 &  190-210 & 88 & 83.8$\pm$7.3 & 0.208$\pm$0.030 & 12730 &48.8 & 56.9 & 0.0061 & 0.0066 \\
220 &  210-230 & 49 & 52.3$\pm$4.7 & 0.214$\pm$0.033 &  7861 &36.5 & 32.2 & 0.0068 & 0.0064 \\
240 &  230-250 & 41 & 37.1$\pm$3.7 & 0.211$\pm$0.038 &  5181 &32.4 & 39.7 & 0.0079 & 0.0087 \\
250 &  240-260 & 34 & 30.1$\pm$3.1 & 0.215$\pm$0.038 &  4417 &28.5 & 35.9 & 0.0080 & 0.0090 \\
270 &  250-290 & 40 & 44.0$\pm$4.5 & 0.297$\pm$0.026 &  2988 &23.5 & 19.6 & 0.0088 & 0.0081 \\
300 &  280-320 & 29 & 26.9$\pm$3.0 & 0.310$\pm$0.029 &  1885 &16.6 & 19.9 & 0.0094 & 0.0102 \\
320 &  300-340 & 22 & 18.3$\pm$2.0 & 0.318$\pm$0.036 &  1371 &13.9 & 18.6 & 0.0100 & 0.0116 \\
350 &  330-370 & 15 & 11.4$\pm$1.2 & 0.311$\pm$0.034 &  902 & 11.2 & 16.3 & 0.0111 & 0.0134 \\
370 &  350-390 & 16 & 8.7$\pm$1.0 & 0.316$\pm$0.039 &  688 &  9.5 & 21.0 & 0.0118 & 0.0175 \\
400 &  380-420 &  7 & 5.8$\pm$0.7 & 0.319$\pm$0.042 &  473 &  7.9 & 10.4 & 0.0129 & 0.0148 \\
450 &  420-480 &  6 & 4.8$\pm$0.6 & 0.366$\pm$0.021 &  259 &  6.7 &  8.2 & 0.0161 & 0.0178 \\
500 &  450-550 &  3 & 5.3$\pm$1.0 & 0.419$\pm$0.014 &  147 &  6.1 &  4.5 & 0.0203 & 0.0175 \\
550 &  500-600 &  1 & 3.3$\pm$0.9 & 0.434$\pm$0.015 &  84.9 & 5.0 &  3.4 & 0.0243 & 0.0200 \\
600 &  550-650 &  1 & 1.84$\pm$0.22 & 0.454$\pm$0.017 & 53.6 & 3.9 &  3.4 & 0.0271 & 0.0251 \\
650 &  600-700 &  2 & 1.04$\pm$0.13 & 0.437$\pm$0.013 & 31.3 & 3.6 &  4.9 & 0.0340 & 0.0396 \\
700 &  620-780 &  2 & 0.84$\pm$0.10 & 0.458$\pm$0.013 & 18.3 & 3.2 &  4.8 & 0.0419 & 0.0513 \\
750 &  660-840 &  1 & 0.51$\pm$0.06 & 0.473$\pm$0.015 & 11.2 & 2.8 &  3.6 & 0.0500 & 0.0573 \\
800 &  700-900 &  1 & 0.32$\pm$0.04 & 0.474$\pm$0.015 &  6.2 & 2.7 &  3.7 & 0.0659 & 0.0775 \\
850 &  750-950 &  0 & 0.18$\pm$0.02 & 0.481$\pm$0.013 &  3.9 & 2.5 &  2.4 & 0.0814 & 0.0799 \\
900 & 790-1010 &  0 & 0.11$\pm$0.02 & 0.475$\pm$0.014 & 2.3 & 2.4 &  2.5 & 0.1030 & 0.1051 \\
950 & 840-1060 &  0 & 0.06$\pm$0.01 & 0.474$\pm$0.012 & 1.3 & 2.4 &  2.5 & 0.1366 & 0.1394 \\
\hline
\hline
\end{tabular}
\end{center}
\end{table*}

\begin{figure}[t]
\includegraphics[width=\figwidth]{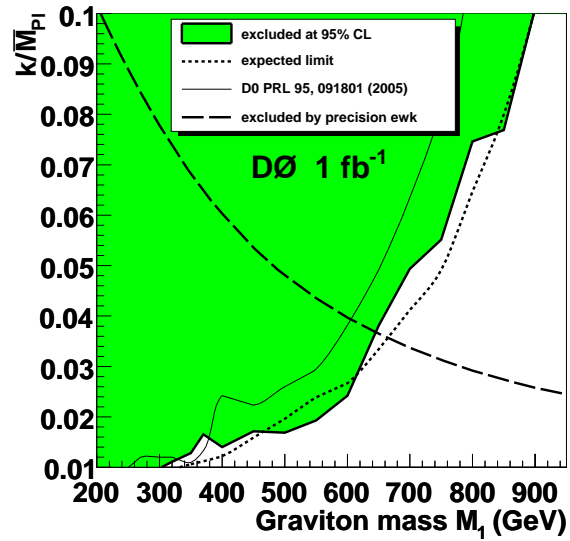}
\caption{95$\%$ confidence level upper limit on $k/\overline{M}_{Pl}$ versus graviton mass $M_{1}$ from $~$1 fb$^{-1}$ of data compared with the expected limit and the previously published exclusion contour~\cite{D0}. The area below the dashed line is excluded by precision electroweak measurements (see~\cite{DHR2}).}
\label{fig:couplinglimit}
\end{figure}

Using the cross section predictions from the Randall-Sundrum model with the same $K$-factor as for the standard model processes~\cite{G}, we set upper limits on the coupling $k/\overline{M}_{Pl}$ as a function of $M_1$. This is shown in Figure~\ref{fig:couplinglimit} and tabulated in Table~\ref{tab:limits}. For $k/\overline{M}_{Pl}=0.01(0.1)$ we can exclude masses below 300(900)~GeV at 95\% confidence level.

In summary, we have searched for RS gravitons as resonances in the $e^+e^-$ and $\gamma\gamma$ invariant mass spectrum from about 1 fb$^{-1}$ of data from the Fermilab Tevatron collider. We find good agreement of the observed spectrum with standard model predictions and set lower limits on the mass of the first massive RS graviton at 95\% confidence level of 300~GeV for $k/\overline{M}_{Pl}=0.01$ and of 900~GeV for $k/\overline{M}_{Pl}=0.1$. These are the tightest direct limits on RS gravitons to date. 

%
We thank the staffs at Fermilab and collaborating institutions, 
and acknowledge support from the 
DOE and NSF (USA);
CEA and CNRS/IN2P3 (France);
FASI, Rosatom and RFBR (Russia);
CAPES, CNPq, FAPERJ, FAPESP and FUNDUNESP (Brazil);
DAE and DST (India);
Colciencias (Colombia);
CONACyT (Mexico);
KRF and KOSEF (Korea);
CONICET and UBACyT (Argentina);
FOM (The Netherlands);
Science and Technology Facilities Council (United Kingdom);
MSMT and GACR (Czech Republic);
CRC Program, CFI, NSERC and WestGrid Project (Canada);
BMBF and DFG (Germany);
SFI (Ireland);
The Swedish Research Council (Sweden);
CAS and CNSF (China);
Alexander von Humboldt Foundation;
and the Marie Curie Program.
%

\end{document}